\documentclass[final,5p,times,twocolumn]{elsarticle}

\usepackage{graphics}
\usepackage{graphicx}
\usepackage{caption}
\usepackage{subcaption}
\usepackage{amssymb}
\usepackage{amsmath}
\usepackage{color}

\DeclareMathOperator{\E}{\mathbb{E}}
\DeclareMathOperator{\R}{\mathbb{R}}
\DeclareMathOperator{\sign}{\mathrm{sign}}
\usepackage{epstopdf}

\biboptions{sort&compress}

\journal{ }

\begin{document}
	
	\begin{frontmatter}
		
		
		
		\title{Market risk factors analysis for an international mining company. Multi-dimensional, heavy-tailed-based modelling.
		}
		
		\author[label1]{{\L}ukasz Bielak}
		\author[label2]{Aleksandra Grzesiek\corref{cor1}}\cortext[cor1]{Corresponding author.}\ead{aleksandra.grzesiek@pwr.edu.wroc.pl}
		\author[label2]{Joanna Janczura}
		\author[label2]{Agnieszka Wy\l oma\'nska}
		\address[label1]{KGHM, M. Sklodowskiej-Curie 48, 59-301 Lubin, Poland}
		\address[label2]{Faculty of Pure and Applied Mathematics, Hugo Steinhaus Center, Wroclaw University of Science and Technology, Wyspianskiego 27, 50-370 Wroclaw, Poland}

		\begin{abstract}
			Mining companies to properly manage their operations and be ready to make business decisions, are required to 
			analyze potential scenarios for main market risk factors. The most important risk factors for  KGHM, one of the biggest companies active in the metals and mining industry, are the price of copper (Cu), traded in US dollars, and the Polish zloty (PLN) exchange rate (USDPLN). {The main scope of the paper is to understand the mid- and long-term dynamics of these two risk factors. For a mining company it might help to properly evaluate potential downside market risk and optimise hedging instruments.}
			From the market risk management perspective, it is also important to analyze the dynamics of these two factors combined with the price of copper in Polish zloty (Cu in PLN), which jointly drive the revenues, cash flows, and financial results of the company. 
			Based on the relation between analyzed risk factors and distribution analysis, we propose to use two-dimensional vector autoregressive (VAR) model with the $\alpha-$stable distribution. The non-homogeneity of the data is reflected in two identified regimes: first - corresponding to the 2008 crisis and second - to the stable market situation.     
			As a natural implication of the model fitted to market assets, we derive the dynamics of the copper price in PLN, which is not a traded asset but is crucial for the KGHM company risk exposure. A comparative study is performed to demonstrate the effect of including dependencies of the assets and the implications of the regime change. 
			Since for various international companies, risk factors are given rather in the national than the market currency, the approach is universal and can be used in different market contexts, like mining or oil companies, but also other commodities involved in the global trading system.

		\end{abstract}
		
		\begin{keyword} metal price \sep exchange rate \sep $\alpha-$stable distribution \sep dependence structure \sep multi-dimensional VAR model \sep regime changes
		\end{keyword}
		
	\end{frontmatter}
	
	
	\section{Introduction}
	Mining companies that produce commodities are exposed to substantial market risk, both on the revenues as well as on the cost side. The same situation is in the case of KGHM Polska Miedz S.A. {(called further KGHM)}, one of the biggest companies active in the metals and mining industry. Among plenty of others, one of the most important risk factors for KGHM are the price of copper (Cu) and Polish zloty (PLN) exchange rate. The first asset is set in global metal exchanges (LME, COMEX, SHFE) in USD, the second one is driven, besides of the relation with global markets, also by the local national economic situation, so their price-setting mechanisms are fundamentally different. However, from the market risk management and modelling perspective, it is important to analyze the dynamics of the behaviour of these two factors combined into the price of copper in Polish zloty (Cu in PLN), which jointly  drives revenues, cash flows, and financial results of the company.
	
	The analysis of the above-mentioned factors, especially the metal price, is a challenging task. One of the reasons is the specific behaviour of the corresponding time series that is manifested by large observations and a non-homogeneous structure. The first mentioned feature, namely, the possible heavy tails of the distribution, implies that the classical Gaussian-based approaches can not be used here. On the other hand, the second-mentioned property, which is the heterogeneous structure, implies that it might be impossible to describe the data with one model having constant (in time) parameters. Hence, we expect that the time series may correspond to different regimes for different time periods. Both features follow directly from the high dependencies of the metal prices, not only on the fundamental factors such as the supply-demand balance but also on the macroeconomic environment, investors' sentiment, and central banks activity \cite{nowa_lukasz1,nowa_lukasz2,nowa_lukasz3}. The above-mentioned factors do not exhaust the list of elements influencing the metal price and the major problem with quantifying their impact is that their magnitude can substantially change over time. It is especially visible in the recent years when many investors make decisions based on information services, headlines and technical signals rather than a thorough fundamental analysis.
	
	From a mining company perspective, the metal price modelling gives the possibility to determine the range of possible price movements, which creates a base for the whole planning process and enables dynamic market risk management using derivatives to shape optimal financial results distributions. In the mining business, the forecasting horizon is in practice middle- and long-term, as daily moves do not determine major business decisions. In that case, the stochastic processes- and time series-based approaches seem to be the most efficient to find potential price behaviour ranges.
	
	Although the methods based on the continuous- and discrete-time processes seem to be appropriate for metal price prediction, one can find many different approaches considered for this problem. The methods proposed in the literature can be in general divided into a few groups: qualitative, trend-based, econometric, stochastic processes- and time series-based. One can also find the modern approach where machine learning techniques are applied  \cite{machine_commodity1,machine_commodity3,machine_commodity4,wang}. Various combinations of these methods are also considered, \cite{prz14,prz16,prz54,prz55,prz56,t2,prz58,prz59,hybrid}. In this paper, we are pointing out only the selected approaches. For more details, we refer the readers, for instance, to the review paper \cite{rs3}. 
	
	In the econometric approach, most of the research is based on the relationship between metal (and generally commodity) prices and economic factors. The dependence between them is the basis for the economic models. This approach is, for instance, demonstrated in \cite{e3,prz67,prz68}, where the similarity between the price movements for different commodities was discussed. The econometric-based methods were also proposed in \cite{e1,e2}. 
	
	In the stochastic-based approach for commodity price modelling, mostly the Gaussian-based models are discussed. The authors argue that predicted market prices based on the stochastic approach are a result of a widely held conviction that market fluctuations have random sources \cite{rs1,rs2,rs3}. The commonly used stochastic models for commodity price modelling are ordinary Brownian motion, geometric Brownian motion, and mean reversion models, {like the Ornstein-Uhlenbeck (OU) process (known as the Vasicek model for interest rate) } or its extension for non-constant coefficients. We refer the readers to the papers \cite{prz92,teor_econ_letter,prz93,prz94,prz95,prz97,jan1}, where different Gaussian stochastic processes are applied for the commodity price modelling. However, some of the authors argue that the Gaussian-based models are inappropriate for commodity price description as they do not take into account the possible large observations apparent in market data \cite{prz97}. Thus, stochastic models based on a more general class of distributions are considered \cite{szarek}. 
	
	The time series-based approach was presented, for instance, in \cite{prz56,t3,t2} where the authors proposed the autoregressive moving average (ARMA) models for a real commodity price.  Moreover, in the papers \cite{prz56,t2,t3} more complicated time series models are applied, like autoregressive conditional heteroscedastic (ARCH), generalized ARCH (GARCH), or autoregressive integrated moving average (ARIMA) models. The error correction model was proposed in \cite{gold1}, while the unequal-interval contour lines and contour time sequences filtration was discussed in \cite{aa2} to predict the metal price. Finally, the improved wavelet–ARIMA time series is presented in \cite{prz59} for commodity price modelling. Similarly to the stochastic processes-based approach, also in time series modelling, the researchers depart from the assumption of Gaussianity and propose the heavy-tailed class of distributions \cite{nowicka}. 
	
	On the border of time series-based and econometric approaches, there are multi-dimensional time series models. On the one hand, multi-dimensional systems can describe the dependence between different factors, and, on the other hand, they take into account the possible dependence in time within one single process. The most classical multi-dimensional time series is the class of vector autoregressive (VAR) models. We refer the readers to the bibliography positions \cite{prze71,prze72,prze70}, where the multi-dimensional modelling of commodity prices and financial markets in the USA and India is considered, respectively. 
	
	Here we also propose the multi-dimensional approach.
	{The \textbf{main scope} of the paper is to understand the mid- and long-term dynamics of the copper price and the USDPLN exchange rate. For a mining company, which is highly dependent on these two risk factors it might help to properly evaluate potential downside market risk and optimise hedging instruments.}
	\textbf{The first objective} of our study is to identify the relationship between movements of the copper price (in USD) and USDPLN exchange rate and to determine its stability over time. Because the data exhibit the non-Gaussian behaviour we express the relationship in the means of different correlation coefficients which are appropriate for heavy-tailed distributed time series. The dependence dynamics analysis  is a base for the multi-dimensional model proposition, what is \textbf{the second goal} of this paper. 
	
	To take into account the possible relationship between the copper price (in USD) and the USDPLN exchange rate, we use the two-dimensional VAR model. However, here we depart from the assumption of the Gaussian distribution of the residual series, {which is shown to be violated,} and use the general class of the $\alpha-$stable distributions {applied} for heavy-tailed distributed data modelling. This class of distributions was initiated by P. L{\'e}vy in 20's \cite{levy,levy1924theorie}. The stable probability laws are important in the probability theory. According to the Generalized Central Limit Theorem, the stable laws attract distributions of sums of random variables
	with a diverging variance. It is a generalization of the Central Limit Theorem, which states that the Gaussian law attracts
	distributions with finite variance. Thus, the $\alpha-$stable distribution is a natural generalization of the Gaussian one. The first application of this distribution appeared in the work of Mandelbrot \cite{mand}, where the financial time series were analyzed. Since that time, the number of research papers devoted to the theoretical and practical aspects related to $\alpha-$stable laws has grown rapidly in different disciplines \cite{shao,nolan:2018}. One can find many interesting applications of the $\alpha-$stable distributions (and in general heavy-tailed ones) and processes in commodity prices and economic factors modelling, see \cite{jan1,MCc,heavy1,heavy2,heavy3,heavy4,heavy5} and references therein. The VAR time series based on the  $\alpha-$stable distribution was considered, for instance, in \cite{grzesiekfloc,ola22,ola11}. 
	{The proposed in this paper model can be considered as the discrete version of the popular multi-dimensional OU process, which has broad applications in finance, e.g., volatility processes in the stochastic volatility models or spread models in the spread options.  The  OU process has the mean-reverting property which is also characteristic for the considered data. The interesting applications of the multi-dimensional OU process with infinite variance one can find also in  \cite{fasen}. The considered in this paper model can be in some sense treated as the discrete version of the model presented in \cite{fasen}.}
	The non-homogeneous behaviour of the data visible in the time series but also in the dependence structure, 
	indicates that there might be a regime change in the market data behaviour. Thus, we utilize the Hidden Markov Model (HMM) approach \cite{rabiner} adjusted for the $\alpha$-stable distribution 
	and divide the data into regimes with a homogeneous structure. {The HMM algorithm considered in this paper takes into account switches in the parameters responsible for the volatility and the heavy-tailed behaviour of the data. The comparable approach was discussed in \cite{sad}, where the author proposed to model the volatility and correlations between emerging market stock prices and the prices of copper, oil and wheal. Obviously, the HMM approach is not the only one that can be used for the structural breaks identification. Among others we refer the readers to the article \cite{mensi}, where structural breaks, dynamic correlation and asymmetric volatility are considered  for petroleum prices and USD exchange rate, or to the paper \cite{rod}, where the relationship between the electric power consumption and economic factors for selected countries are discussed also in the context of the structural breaks.} 
	
	\textbf{The third objective} of the paper is a comparative analysis of the analyzed data distribution under different market conditions, reflected in the regime change. The multi-dimensional data modelling, involving the regime-changing behaviour, gives the possibility to analyze the copper price in USD together with the USDPLN exchange rate. Importantly, it is also a baseline for the copper price in PLN analysis, which is a product of the former two variables.  Here, two approaches are compared. Both of them are based on the heavy-tailed VAR system. However, in the first approach, the relationship between the considered assets is taken into account while in the second one - it is assumed to be negligible. The comparative study of these two approaches is \textbf{the fourth objective} of the paper. Since commodity price in the national currency is the main risk factor for different international companies, the proposed approach can be utilized in various market contexts.
	

	The rest of the paper is organized as follows. In Section \ref{problem} we formulate the problem. Next, in Section \ref{methodology} we present the used methodology, namely, we describe the applied dependency measures, the $\alpha-$stable distribution, and the $\alpha-$stable distributed VAR model. Then, in Section \ref{multi} we analyze the two-dimensional real data sets describing the copper price (in USD) and USDPLN exchange rates using the $\alpha-$stable VAR time series {and show its advantages over the standard Gaussian VAR model}. 
	In the last section, we discuss the results and conclude the paper.

	\section{Problem formulation}\label{problem}
	{The main problem considered in this paper  is to understand the  dynamics of the copper price (in USD) and the USDPLN exchange rate and also the dynamics of these two factors combined into the price of copper in Polish zloty (Cu in PLN).} Although the formulated problem seems to be dedicated to the specific case, it can be considered as a general one, namely, the prediction of the range of values for the metal price expressed in the currency of a given country, when the relationship between the metal price and the exchange rate (between the original currency and the national currency) is changing in time. The non-Gaussian behaviour of both assets makes the problem more difficult. From a market risk management perspective, obtaining reasonable extreme observation levels (optimistic and pessimistic) is a crucial objective. These levels should be wide enough to capture market movements and at the same time narrow adequately to enable making business decisions. It is important to capture in the modelling of the copper price in PLN the non-Gaussian characteristics, changing regimes, and non-constant relation between risk factors, as they create additional market risk for the company, if they occur. Such modelling can be used for the assessment of the company market risk and the verification of whether tools like, for example, derivatives properly mitigate the excess and unacceptable risk of the company.
	\begin{figure}[ht!]
		\begin{center}
			\includegraphics[scale=0.6]{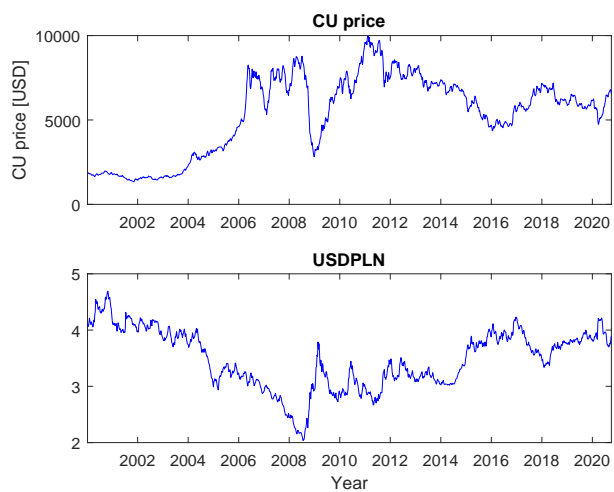}
			\caption{The weekly data corresponding to the copper price in USD (top panel) and USDPLN exchange rate (bottom panel) from the period Jan 7th, 2000 - October 2nd, 2020. }\label{fig2}
		\end{center}
	\end{figure}
	\begin{figure}[ht!]
		\begin{center}
			\includegraphics[scale=0.6]{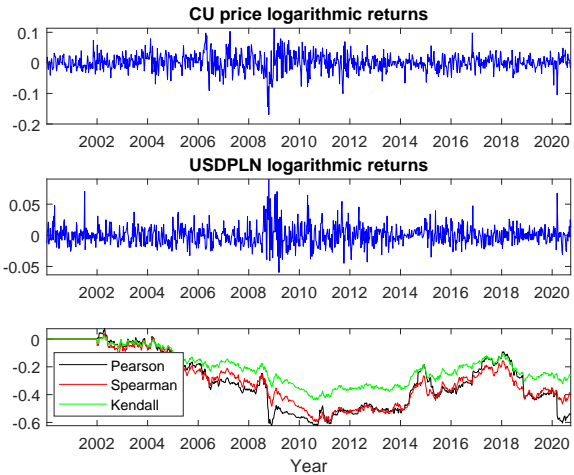}
			\caption{The logarithmic returns of the weekly copper price in USD (top panel), USDPLN exchange rate (middle panel), and the corresponding sample Pearson, Spearman rank, and Kendall rank correlation coefficients calculated for two-yearly windows ($104$ observations).}\label{fig3}
		\end{center}
	\end{figure}

	The analyzed data sets are shown in Fig. \ref{fig2}. They correspond to the weekly copper prices in USD (top panel) and USDPLN exchange rate (bottom panel) from the period January 7th, 2000 - October 2nd, 2020. {In this paper, we examine the logarithmic returns of the analysed assets. The analysis of logarithmic returns instead of raw time series allows to get rid of the influence of deterministic trends that may obscure important information hidden in the data such as heavy-tailed behavior or regime changes. This approach  is  practiced in the company for the risk management purpose as it takes into account the volatility changes and the heavy-tailed behaviour.  These properties mostly influence the confidence intervals of the future prices or exchange rates. } The logarithmic returns of the analyzed data are demonstrated in Fig. \ref{fig3} in the top and middle panels, respectively. One can see the specific behaviour of the time series. On one hand, it is clearly seen that both data sets exhibit a non-Gaussian behaviour with visible large observations that may suggest a heavy-tailed distribution of the time series. On the other hand, one can see that the data have a non-homogeneous structure - there are visible regimes of the time series that do not fit the overall pattern. Special attention should be paid to the period from the year 2008 to 2012, where large observations are more frequent than in the other periods. This behaviour is especially visible for the copper price logarithmic returns (see the top panel of Fig. \ref{fig3}). The non-homogeneity of the time series indicates that the parameters switch in time and one stationary model can not be used for the whole data description.
	
	Despite the fact that there is no direct economic relationship between the copper price (in USD) and the USDPLN exchange rate, even a short look at the charts shows that there is a negative relation between these risk factors, see Fig. \ref{fig3}, the bottom panel. From the market risk management point of view, such statement is essential, especially if these assets are the major risk factors in the portfolio. However, if that relation would turn out to be apparent or unstable during time, it could lead to false conclusions and in consequence wrong decisions. 
	
	The mentioned relation between the considered assets can be expressed by the means of different measures of dependence. In this paper, we use three metrics, namely, the Pearson, Spearman rank, and Kendall rank correlation coefficients. The definitions and properties of these measures 
	are presented in the next section. Here we only mention that the Pearson correlation is effective for the data with light-tailed distribution (like the Gaussian one), while the two other correlation measures can be used also for the heavy-tailed distributed time series. To demonstrate the dynamics of the dependence structure, the sample correlation coefficients are calculated for the data from a moving window corresponding to a two-yearly period ($104$ observations). 
	As one can see in Fig. \ref{fig3} {(the bottom panel)}, the sample correlation coefficients change over time, which indicates the dynamical structure of the relationship between the assets. Moreover, for the period between the years 2006 and 2012, there is a clear difference between the Pearson correlation coefficient and the two other measures (especially the Kendall rank correlation coefficient). This phenomenon may confirm the non-homogeneous structure of the data. Moreover, the heavy-tailed distribution may also influence the difference between the dependency measures.

	The specific characteristics of the data described above are the motivation for using non-Gaussian models for their description. Moreover, the visible relationship between the analyzed assets implies that a multi-dimensional model needs to be applied. The dynamics of the correlation coefficients and the non-homogeneous behaviour of the data indicate that in the first step of the analysis the examined time series should be divided into regimes of homogeneous structure.  
	
	\section{Methodology}\label{methodology}
	In this section, we present the general methodology used in the analysis. The demonstrated methods are known from the literature, thus we recall only the main concepts, definitions, and properties.  
	\subsection{The dependence structure description for Gaussian and non-Gaussian time series}
	Different measures of dependence between variables may be used to describe the interplay of elements in a complex system and the strength of their relationship. In this paper, we focus on three most broadly used measures, namely, the Pearson correlation, the Spearman rank correlation, and the Kendall rank correlation \cite{eurphys}{, and use them for the illustration of the dependence characteristics.}
	
	The Pearson correlation coefficient $\rho_P$ for a random vector $(X,Y)$ is defined as \cite{dunn2009basic}:
	\begin{equation}
	\rho_{P} =\frac{\text{cov}(X,Y)}{\sigma_X\sigma_Y},
	\label{pearson1}
	\end{equation}
	where $\text{cov}(\cdot,\cdot)$ is the covariance function, $\sigma _{X}$ is the standard deviation of $X$ and $\sigma_Y$ is the standard deviation of $Y$. 
	{
		It is used to study the linear relationship between two variables 
		and is sensitive to outliers. Therefore, this measure is useful especially for the Gaussian (or light-tailed) distributed variables.}
	
	The Spearman rank correlation coefficient for a random vector $(X,Y)$ has the following form \cite{kendall1948rank,kend2}:
	\begin{equation}
	\rho_{S} =\frac{\text{cov}(Q,W)}{\sigma_{Q}\sigma_{W}},
	\label{spearman1}
	\end{equation}
	where $(Q,W)$ is a random vector of ranks corresponding to $(X,Y)$,  $\sigma _{Q}$ and $\sigma _{W}$ are the standard deviations of variables $Q$ and $W$, respectively. 
	{The Spearman rank correlation 
		measures a monotonic relationship. It is insensitive to large observations and thus in cases when the analyzed variables are heavy-tailed, describes the relation more adequately than the Pearson correlation.}
	
	The last considered measure is the Kendall rank correlation. Let  $(x_1,y_1), (x_2,y_2), \dots, (x_n,y_n)$ be a random sample corresponding to the random vectors $(X, Y)$. The sample Kendall rank correlation coefficient is defined as follows \cite{kendall1938new}:
	\begin{equation}
	r_K =\frac{ 2 }{ n(n-1) } \sum_{1 \leq i \leq j \leq n} J((x_i,y_i),(x_j,y_j)),
	\label{kendal}
	\end{equation}
	where $J((x_i,y_i),(x_j,y_j)) =\text{sgn}(x_i-y_i) \text{sgn}(x_j-y_j)$ and $J((x_i,y_i),(x_j,y_j))= 1,$ if a pair $(x_i,y_i)$ is concordant with a pair $(x_j,y_j),$ i.e. if $(x_i-x_j)(y_i-y_j) > 0; J((x_i,y_i),(x_j,y_j)) =-1,$ if a pair $(x_i,y_i)$ is discordant with a pair $(x_j,y_j),$ i.e. if $(x_i-x_j)(y_i-y_j) < 0.$
	The Kendall rank correlation coefficient is based on the difference between the probability that two variables are in the same order (for the observed data vector) and the probability that their order is different. In formula (\ref{kendal}) it is required that the variable values can be ordered. 
	{
		The Kendall rank correlation coefficient indicates not only the strength but also the direction of the dependence. Similarly to the Spearman rank correlation, it is resistant to outliers and is used especially for the non-Gaussian distributed data \cite{eurphys}. Let us mention, that, besides the mentioned correlation coefficients, other dependency measures adequate for heavy-tailed distributed data are also considered in the literature, see for instance \cite{dedi3,wylomanska2015codifference,ma1996joint}. }
	
	\subsection{The $\alpha-$stable distribution}
	
	The data analyzed in this paper are non-Gaussian. To model such a specific behaviour, we propose to apply the $\alpha-$stable distribution. Below we recall the corresponding definition and the main properties. 
	
	For the $\alpha-$stable distribution the probability density function (PDF) and the cumulative distribution function (CDF) are given in close form only in a few special cases. Therefore, a common way to define the distribution of an $\alpha-$stable random variable ${Z}$ is by determining its characteristic function \cite{stable}:
	\begin{align} \label{eq:stable_cf}
	\E \left[ \exp\{i\theta Z\} \right] =
	\exp\left\{-\sigma^\alpha|\theta|^\alpha\left(1+i\beta w(t,\theta)\right)+i\mu\theta)\right\},
	\end{align}
	where:
	\begin{align}
	w(\theta,\alpha) = \left\{ \begin{array}{ll}
	-\sign\left(\theta\right)\tan\left(\frac{\pi\alpha}{2}\right) & \textrm{if $\alpha\neq 1$,}\\
	\frac{2}{\pi}\sign\left(\theta\right)\ln\left|\theta\right| & \textrm{if $\alpha= 1$,}
	\end{array} \right. 
	\end{align}
	and $\sign(\cdot)$ denotes a sign function. The parameter $0<\alpha\leq 2$ is called the stability index and regulates the rate at which the distribution tails converge. The other parameters are: the scale parameter $\sigma > 0 $, the skewness parameter $-1\leq\beta\leq1$, and the shift parameter $\mu \in \R$. If $\beta=\mu=0$, the distribution of $Z$ is symmetric with respect to $0$ and the characteristic function given in Eq. (\ref{eq:stable_cf}) simplifies to the following one:
	\begin{align} \label{eq:stable_cf_symmetric}
	\E \left[ \exp\{i\theta Z\} \right] =
	\exp\left\{-\sigma^\alpha|\theta|^\alpha\right\}.
	\end{align}
	
	It is worth emphasizing that the $\alpha-$stable distribution with $0<\alpha<2$ constitutes a generalization of the Gaussian distribution corresponding to the case of $\alpha=2$. For the non-Gaussian distribution, the properties differ significantly from the ones corresponding to $\alpha=2$: the tails converge to zero according to a power function and the second moment is infinite. Additionally, for $0<\alpha\leq1$ the first moment is also infinite. As a consequence, the  $\alpha-$stable random variables take extreme values more likely than it is observed in the Gaussian case. 
	
	For more information about the one-dimensional as well as the multi-dimensional $\alpha-$stable distribution we refer the readers to \cite{PRESS_multi,PAULAUSKAS,Weron_survey,zolotarev,janicki_weron,stable}.

	\subsection{Vector autoregressive model with $\alpha-$stable distribution}
	
	Vector autoregressive time series (also called the VAR model) with the $\alpha$-stable distribution is defined as an extension of the classical model where the innovations are assumed to be Gaussian distributed (or at least have finite second moments), see for example \cite{brockwell2016introduction}. In the $\alpha$-stable non-Gaussian case with infinite variance, the VAR system can be used to model the data exhibiting a higher likelihood of more extreme events.
	
	A time series $\{\mathbf{X}(t)\}= \{(X_1(t),\ldots,X_m(t))^T\}$ is called a vector autoregressive model with the $\alpha$-stable distribution, if for each $t \in \mathbb{Z}$ it satisfies the following system of equations:
	\begin{align} \label{eq:autoregressive}
	\mathbf{X}(t)-\Theta_1\mathbf{X}(t-1)-\ldots-\Theta_p\mathbf{X}(t-p)=\mathbf{Z}(t),
	\end{align}
	where $\{\mathbf{Z}(t)\}=\{(Z_1(t),\ldots,Z_m(t))^T\}$ is a $m$-dimensional $\alpha-$stable random vector and $\Theta_1,\ldots,\Theta_p$ are $m\times m$ matrices with time-constant coefficients. For simplicity, we assume here that the noise vector $\mathbf{Z}(t)$ consists of independent $\alpha$-stable distributed random variables, i.e. $Z_i(t)$ and $Z_j(t)$ are independent for any $t\in \mathbb{Z}$ when $i\neq j$, and the characteristic function of $Z_i(t)$ is given by Eq. (\ref{eq:stable_cf}) for all $t\in\mathbb{Z}$. Moreover, the vector $\mathbf{Z}(t)$ is assumed to be independent of the vector $\mathbf{Z}(s)$ for $t\neq s$, where $t,s\in \mathbb{Z}$.
	
	The conditions for the existence and uniqueness of the bounded solution of the vector autoregressive time series with the multi-dimensional $\alpha-$stable distribution are provided in \cite{PeirisThavaneswaran}. Note that for $p=1$, the bounded solution of the autoregressive system of order $1$ takes the form:
	\begin{align}
	\mathbf{X}(t)=\sum_{j=0}^{+\infty}\Theta^j\mathbf{Z}(t-j),
	\end{align}
	under the assumption that the elements of $\Theta^j$ are absolutely summable, i.e., if the eigenvalues of $\Theta$ are less than $1$ in the absolute value, where $\Theta=\Theta_1$ in Eq. (\ref{eq:autoregressive}). It should be mentioned that in the case when the coefficients of the matrices in Eq. (\ref{eq:autoregressive}) responsible for the relationship between time series components are zero, then the VAR model reduces to $m$ independent one-dimensional $\alpha-$stable autoregressive (AR) time series \cite{nw1}. 
	
	In the classical (Gaussian) version of the VAR system, the dependence structure of the process can be described using the covariance or the correlation. As a consequence, to estimate the parameters of the system, one often uses the multi-dimensional Yule-Walker method based on the auto-covariance function \cite{brockwell2016introduction}. However, since for the VAR model with non-Gaussian $\alpha-$stable distribution the second moment is infinite, there is no theoretical justification for using the covariance-based method to estimate the unknown parameters. Therefore, in \cite{covariationYule-Walker} the authors propose the modified Yule-Walker method, similarly to the one-dimensional case examined in \cite{est2,physica}, which is based on the covariation well defined for the $\alpha-$stable distribution with $\alpha>1$.  The covariation can be also used, instead of the auto-covariance, to quantify the interdependence within a time series. In this case, the measure is called auto-covariation and it is defined in the following way:
	\begin{equation}
	\mathrm{{CV}}(X_i(t),X_i(t-h))=\frac{\mathbb{E}[X_i(t)X_i(t-h)^{\langle p-1 \rangle}]}{\mathbb{E}[|X_i(t-h)|^p]}\sigma_{X_i(t-h)}^\alpha,
	\end{equation}
	where:
	\begin{equation}
	x^{\langle a \rangle}=|x|^{a}\mathrm{sign}(x),
	\end{equation}
	$1<p<\alpha$ and $\sigma_{X_i(t-h)}$ is the scale parameter of the random variable ${X_i(t-h)}$. In practice, one often estimates the so-called normalized auto-covariation from the data, i.e., the auto-covariation divided by the parameter $\sigma_{X_i(t-h)}$. The appropriate estimators are presented in \cite{nikias1995,est2,physica}.
	
	\section{Two-dimensional analysis of the copper price (in USD) and USDPLN exchange rate}\label{multi}
\begin{figure}[ht!]
		\begin{center}
			\includegraphics[scale=0.6]{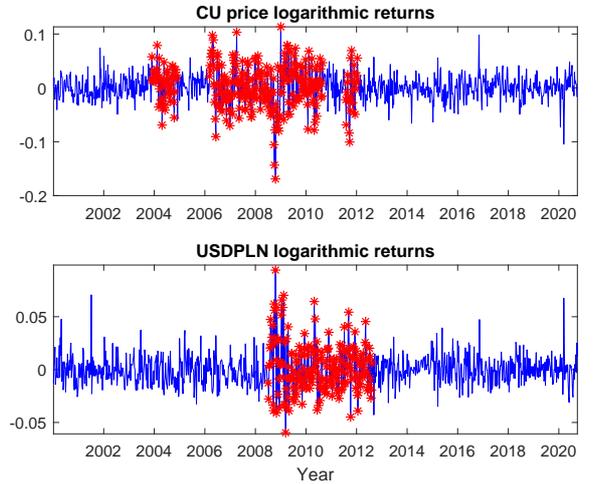}
			\caption{The logarithmic returns of the weekly copper price in USD (top panel) and  USDPLN exchange rate (bottom panel) with regimes obtained via the HMM classification procedure.}\label{fig4}
		\end{center}
	\end{figure}

	In this section, we present the analysis of two-dimensional modelling for data corresponding to the copper price (in USD) and the USDPLN exchange rate. The visual inspection of the logarithmic returns (see Fig. \ref{fig3} top and middle panels) of the considered assets and the evident difference between the correlation coefficients (see Fig. \ref{fig3}, bottom panel) indicate that the data are related, however, they should be divided into parts of the homogeneous structure. In order to do this, we assume that the one-dimensional time series of logarithmic returns follow a symmetric $\alpha-$stable distribution with parameters $\sigma$ and $\alpha$ switching between two values. These two parameter sets $(\sigma_1, \alpha_1)$ or $(\sigma_2, \alpha_2)$ are corresponding to an unobserved state process and, hence, reflect the changes in the market conditions. To estimate the moments of switching, we apply a HMM approach \cite{rabiner} and assume that the state process is driven by a Markov chain with probabilities of changing the states given by a transition matrix.  {The pre-processing step of HMM classification,  aims at distinguishing phases with low and high variations, reflecting the uncertainty of the analyzed risk factors. These variations are jointly modelled by the scale parameter $\sigma$ and the stability parameter $\alpha$, describing heaviness of the distribution tails.}  The HMM estimation procedure is based on the expectation-maximization algorithm, \cite{dempster}, designed to infer parameters in the models depending on latent variables (here the state process). As a by-product of the EM algorithm, we obtain the probabilities of the two states for each time point. These probabilities are then used for the identification of different regimes within the time series. Namely, for each value of the logarithmic returns, we assign the state that is more probable. 
	The resulting regime classification for both variables is illustrated in Fig. \ref{fig4}.
	{It indicates on the existence of distinct regimes with low/high variations given by both parameters.}
	The obtained results are consistent with the market situation reflected in the analyzed data. Few years before and after the peak of the great financial crisis (2008) the volatility in the market among many different assets stayed at elevated levels, whereas in other periods the market moves were significantly weaker. One of the reasons for this issue could be that before and after the crisis, valuations of many assets have achieved extreme levels, with very dynamic changes also reflected in the currency markets. In the case of the copper price the data showed that the elevated volatility has been observed even earlier as a result of the substantial incremental Chinese growth dynamics. For further analysis, we decided to choose the overlapping regimes timing for both assets. 
	{For the verification of the regimes identification we have also applied the ICSS algorithm of \cite{ICSS} designed for detecting structural breaks in variance. The obtained results were corresponding the the HMM results. However, the HMM model uses exactly the $\alpha$-stable distribution assumed in the analyzed model, so it is better suited for detecting changes of the parameters in the considered case. A similar market regimes classification was also found for a portfolio of stock prices in \cite{poela}, where the authors incorporated the switching mechanism into the correlation matrix instead of the distribution parameters. This shows that the general market behaviour change was reflected both in the correlations and in the distributions. Indeed, for the illustration related to the considered assets, we have also calculated the Spearman and Kendall correlation coefficients corresponding to the identified regimes using a four-yearly moving window. In order to keep the time relation with the returns, the symmetric windows were used, i.e. for the calculation of the  correlation coefficient for a given time point, the data from the preceding and the succeeding two years were used. The obtained results are plotted in the bottom panel of Fig. \ref{fig:corr_reg}. Note that the windows containing observations from both regimes were omitted, so the correlation curves have gaps around the regime change. Clearly, we can observe a change in the correlation coefficients level related to the regime change. }

	\subsection{The $\alpha-$stable VAR modelling involving relationship between the considered assets}
	
\begin{figure}[ht!]
		\begin{center}
			\includegraphics[scale=0.6]{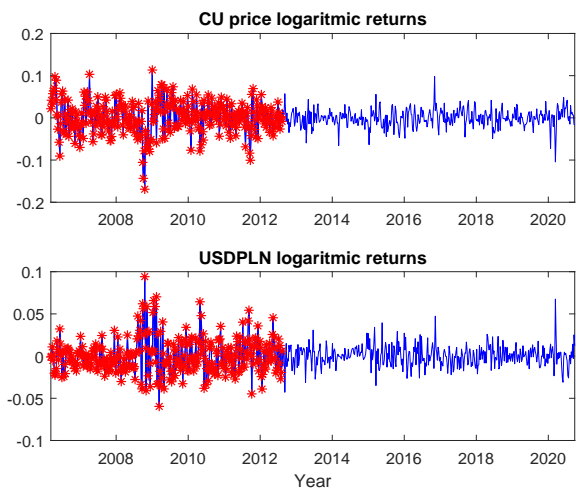}
			\caption{The logarithmic returns of the weekly copper price in USD (top panel) and USDPLN exchange rate (bottom panel) with marked regimes used for time series modelling. Regime 1 is marked with red stars, while regime 2 is marked in blue. }\label{fig5}
		\end{center}
	\end{figure}

\begin{figure}
		\centering
		\includegraphics[scale=0.6]{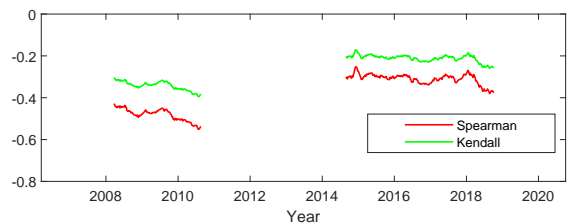}
		\caption{{The Spearman and Kendall correlation coefficients corresponding to the identified regimes (see Fig. \ref{fig5}). They were calculated using a symmetric four-yearly window. The windows containing observations from both regimes were omitted.}}
		\label{fig:corr_reg}
	\end{figure}

	In this section, we model the logarithmic returns of the copper price (in USD) and the USDPLN exchange rate using the two-dimensional VAR time series with the $\alpha$-stable distribution, described in Section \ref{methodology} {and compare its fit with the standard Gaussian VAR model}. We assume the simplest version of the model, namely VAR(1). In this approach, we allow for a possible dependence between the considered assets, in contrast to the second approach presented in the next subsection. Taking into account the regime identification step, we assume that the parameters of the VAR(1) model change at a certain point in time. Therefore, we separately consider regime 1 and regime 2 marked in Fig. \ref{fig5}.   In our analysis, we assume that regime 1 starts when the first of the assets (copper price in USD or USDPLN exchange rate) falls into this regime due to the HMM classification step (see  Fig. \ref{fig4}). Note that we omit the short period in 2004, where regime 1 was identified for the copper price, since there is no corresponding regime change in the USDPLN exchange rate. Thus, regime 1 starts in March of 2006. In the economical context of world exchanges, including commodities markets, there was a dynamic growth of assets value starting from 2006, which led eventually to a financial crisis outbreak two years later. 
	The end of regime 1 is specified as the second half of 2012 when the situation on the market has started to stabilize and the classification results indicate  the second regime for both assets. 
	The final regimes segmentation is plotted in Fig. \ref{fig5}. In practice, we separately fit the VAR(1) model for each regime. 
	
\begin{figure}[ht!]
		\begin{center}
			\includegraphics[scale=0.6]{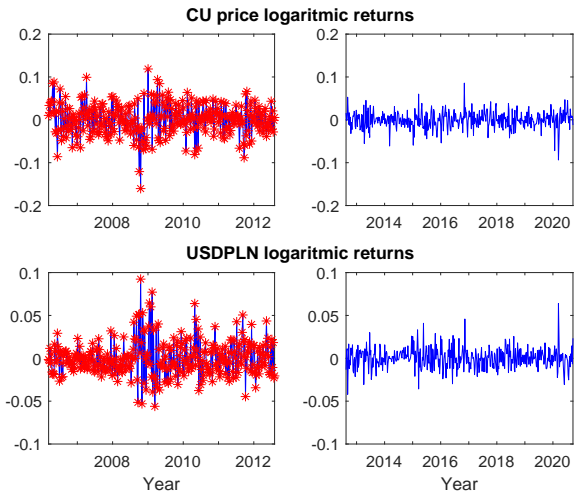}
			\caption{The residuals corresponding to the VAR(1) model applied to the logarithmic returns of the weekly copper price in USD and USDPLN exchange rate for the data from regime 1 (left panels) and regime 2 (right panels).}\label{fig6}
		\end{center}
	\end{figure}

	The matrices of the coefficients of the two-dimensional VAR(1) models estimated based on the data corresponding to regime 1 and regime 2, respectively, are given by: 
	\begin{align}
	&\widehat{\Theta}_{regime1}=\left[
	\begin{array}{cc}
	0.2706 & -0.0569\\
	0.0063 & \ \ \ 0.2134
	\end{array}
	\right],\nonumber\\ 
	&\widehat{\Theta}_{regime2}=\left[
	\begin{array}{cc}
	0.3119 & \ \ \ 0.1403\\
	0.0010 & \ \ \ 0.1472
	\end{array}
	\right]
	\label{eq:Theta_Segments}.
	\end{align}
	The estimation results indicate that there exists a relation between the considered factors in both regimes.  The parameters related to the dependence between the assets lie outside the main diagonal and have non-zero values. The corresponding residual time series are presented in Fig. \ref{fig6}. We recall that in the VAR(1) model the residual vectors are assumed to be independent and identically distributed. This holds also in the one-dimensional sense, i.e., for the components of the residual vectors treated separately. In Fig. \ref{fig_autocv_two} given in the Appendix, we plot the corresponding auto-covariation functions which indicate a non-zero value only for $h=0$. Recall that the auto-covariation function, similarly to the auto-covariance function for the Gaussian (or generally light-tailed) case, corresponds to the interdependence of time series.
	
	To demonstrate that the residuals of the model are not Gaussian distributed, we use five goodness-of-fit tests based on the distances between the empirical and theoretical cumulative distribution functions (CDF). The empirical CDF is calculated for the residual series, while the theoretical one is the CDF of the Gaussian distribution with parameters estimated from the residual series. Here we use the following statistical tests: Kolmogorov-Smirnov test (T1) \cite{ks}, Kuiper test (T2) \cite{kuiper}, Watson test (T3) \cite{watson}, Cramer-von Mises test (T4) \cite{cvm} and Anderson-Darling test (T5) \cite{ad}. In Table  \ref{tab:table1} we present the obtained p-values. As one can see, the $p$-values for $H_0$ hypothesis of the Gaussian distribution are relatively small, so Gaussianity can be rejected for most of the considered cases, except for the CU price in the first regime, at the standard $0.05$ significance level.  
	This result is a motivation for the $\alpha-$stable distribution testing. We use the same  T1-T5 goodness-of-fit tests, however, the theoretical CDF is calculated for the $\alpha-$stable distribution with the parameters fitted to the corresponding residual series, see Table \ref{tab:table1} for the results of the tests. All considered tests indicate that there is no evidence against the null hypothesis that the residuals are $\alpha$-stable distributed (all of the obtained p-values are higher than the standard significance level of $0.05$). 
	{To compare the $\alpha$-stable VAR model with the standard Gaussian one, we also calculate the coverage rates at 0.05, 0.1, 0.25, 0.75, 0.9 and 0.95 levels. The coverage rates are calculated as the percentage of observations below a given quantile of the model distribution. For a data following an assumed model the coverage rates should be equal to the quantile level. The obtained results are given in Table \ref{tab:table1}. In the Gaussian case the coverage rates are close to the expected ones only for the cooper price  in the first regime, what corresponds to the goodness-of-fit results. For all other cases the Gaussian VAR(1) model yields underestimated low quantiles and overestimated high quantiles. This is not the case for the $\alpha$-stable VAR(1) model, since most of the values are close the expected rates.}
	
	In Table \ref{tab:table2} we present the results of fitting the one-dimensional $\alpha$-stable distribution to each residual time series separately. For the estimation of the $\alpha-$stable distribution parameters we applied the regression method \cite{reg}. Note that for both assets we obtained the higher $\sigma$ values in the first regime, with the value being almost two times larger than in the second regime. It shows that there was a significant change in the scale of the market fluctuations between the regimes. 
	
	
	\begin{table*}[htbp] 
		\centering 
		\caption{{The $p$-values of the goodness-of-fit tests (T1: Kolmogorov-Smirnov, T2: Kuiper, T3: Watson, T4: Cramer-von Mises, T5: Anderson-Darling) for the residuals distribution and the coverage rates obtained for the two-dimensional Gaussian as well as $\alpha$-stable VAR(1) models.}}\vspace{0.1cm}
		\begin{tabular}{|c|c|c|c|c|c|c|c|c|c|c|c|} \hline
			Model & \multicolumn{5}{|c|}{Goodness-of-fit test $p$-value}&\multicolumn{6}{|c|}{Coverage rate}\\ \hline
			&T1  & T2 & T3 & T4 & T5& 0.05 & 0.1 & 0.25 & 0.75 & 0.9& 0.95\\
			\hline\hline 
			\multicolumn{12}{|c|}{CU Price REGIME 1} \\ \hline 
			Gaussian	&	$0.1580$ & $0.0730$ & $0.0530$ & $0.0460$ & $0.0590$ &0.0478&	0.0955&	0.2119&	0.7582&	0.9075&	0.9552
			\\\hline 
			$\alpha$-stable	&		$0.4080$ & $0.3230$ & $0.2980$ & $0.2260$ & $0.3930$&0.0507&	0.0985&	0.2299&	0.7493&	0.8955&	0.9433
			\\\hline \hline
			\multicolumn{12}{|c|}{USDPLN REGIME 1} \\ \hline 
			Gaussian	&		$0.0000$ & $0.0000$ & $0.0000$ & $0.0000$ & $0.0000$&0.0269&	0.0627&	0.2328&	0.7940&	0.9224	&0.9403
			\\\hline 
			$\alpha$-stable&		$0.5440$ & $0.7760$ & $0.8850$ & $0.9520$ & $0.4600$ &0.0627&	0.1015&	0.2537&	0.7672&	0.9134&	0.9493
			\\\hline \hline
			\multicolumn{12}{|c|}{CU Price REGIME 2} \\ \hline 
			Gaussian	&		$0.0380$ & $0.0210$ & $0.0030$ & $0.0020$ & $0.0010$&0.0449&	0.0757&	0.2104&	0.7943&	0.9078&	0.9622
			\\\hline 
			$\alpha$-stable&		$0.9090$ & $0.9550$ & $0.9910$ & $0.9730$ & $0.9680$ & 0.0520&	0.0946&	0.2411&	0.7730&	0.8936&	0.9338
			\\\hline \hline
			\multicolumn{12}{|c|}{USDPLN REGIME 2} \\ \hline 
			Gaussian	&		$0.0010$ & $0.0010$ & $0.0020$ & $0.0010$ & $0.0010$ &0.0449&	0.0875&	0.2222&	0.7801&	0.9243&	0.9504
			\\\hline 
			$\alpha$-stable&	$0.7880$ & $0.6650$ & $0.4680$ & $0.4780$ & $0.3560$ &0.0473	&0.1017&	0.2364	&0.7400	&0.9078	&0.9456
			\\\hline 
		\end{tabular}
		\label{tab:table1} 
	\end{table*}
	\begin{table}[htbp] 
		\centering 
		\caption{The parameters of the $\alpha$-stable distribution estimated for the residual time series corresponding to the two-dimensional VAR(1) models.}\vspace{0.1cm}
		\begin{tabular}{|c|c|c|c|} \hline
			$\alpha$  & $\sigma$ & $\beta$ & $\mu$\\
			\hline\hline 
			\multicolumn{4}{|c|}{CU Price REGIME 1} \\ \hline 
			$1.9219$ & $0.0236$ & $-0.5714$ & $0.0007$ \\ \hline 
			\multicolumn{4}{|c|}{USDPLN REGIME 1} \\ \hline 
			$1.7229$ & $0.0114$ & $1.0000$ & $0.0016$ \\ \hline 
			\multicolumn{4}{|c|}{CU Price REGIME 2} \\ \hline 
			$1.8243$ & $0.0119$ & $-0.3416$ & $-0.0004$ \\ \hline 
			\multicolumn{4}{|c|}{USDPLN REGIME 2} \\ \hline 
			$1.8424$ & $0.0074$ & $-0.0160$ & $0.0000$\\\hline
		\end{tabular}
		\label{tab:table2} 
	\end{table}
	Using two-dimensional $\alpha-$stable VAR(1) model is consistent with the economic reality, where the relation between market factors apparently unrelated like, which are in our case the copper price (in USD) and the USDPLN exchange rate, is visible and even growing over time in recent times, due to the large amounts of money put by central banks into the circulation.
	
	
	\subsection{The $\alpha-$stable VAR modelling involving no relationship between the considered assets}
	In this part, we present the results obtained under the assumption that the relation between the copper price (in USD) and USDPLN exchange rate is negligible and can be omitted. Thus, in this approach it is assumed that in the two-dimensional $\alpha-$stable VAR(1) model the coefficients outside the main diagonal are equal to zero, so the components are independent. Similarly to the previous case, we fit two models to the data, separately for regime 1 and regime 2, chosen in the same manner as previously.
	The coefficients of the models in the second considered approach are as follows: 
	\begin{align}
	&\widehat{\Theta}_{regime1}=\left[
	\begin{array}{cc}
	0.2927 & 0\\
	0 & \ \ \ 0.2100
	\end{array}
	\right],\nonumber\\ 
	&\widehat{\Theta}_{regime2}=\left[
	\begin{array}{cc}
	0.2810 & \ \ \ 0\\
	0 & \ \ \ 0.1386
	\end{array}
	\right]
	\label{eq:Theta_Segments_onedim}
	\end{align}
	and the residual time series are presented in Fig. \ref{fig13}. In Fig. \ref{fig_autocv_one} given in the  Appendix we additionally plot the corresponding auto-covariation functions of the residuals having non-zero values only for $h=0$. Similarly to the first approach, in Table \ref{tab:table3} we present the results for the Gaussian and $\alpha-$stable distribution testing for the residual series. All tests show no evidence for  rejecting the hypothesis about the $\alpha$-stable distribution. On the other hand, the p-values for the Gaussian distribution testing are significantly smaller than the $5\%$ significance level in all cases except for the CU price in the first regime. {The coverage rates obtained for both models also show that in terms of fit the $\alpha$-stable distribution outperforms the Gaussian one.}
	
	In Table \ref{tab:table4} we present the estimated parameters of the $\alpha$-stable distribution for the residual time series. 
	\begin{figure}[ht!]
		\begin{center}
			\includegraphics[scale=0.6]{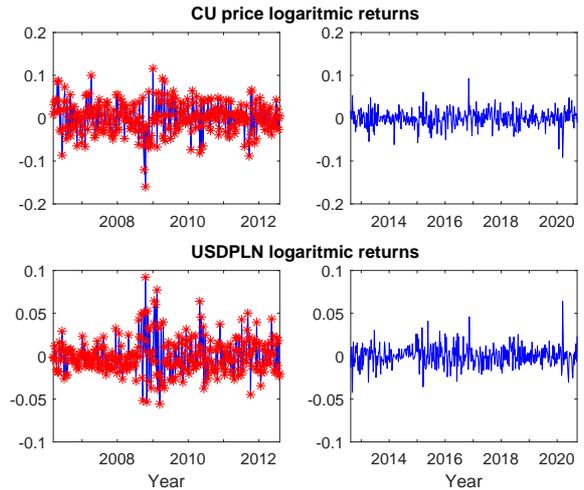}
			\caption{{The residuals corresponding to the VAR(1) model with independent components applied to the logarithmic returns of the weekly copper price in USD and USDPLN  exchange rate for the data from regime 1 (left panel) and regime 2 (right panel).}}
			\label{fig13}
		\end{center}
	\end{figure}
	
	\begin{table*}[htbp] 
		\centering 
		\caption{{The $p$-values of the goodness-of-fit tests (T1: Kolmogorov-Smirnov, T2: Kuiper, T3: Watson, T4: Cramer-von Mises, T5: Anderson-Darling) for the residuals distribution and the coverage rates obtained for the Gaussian as well as $\alpha$-stable VAR(1) models with independent components.}}\vspace{0.1cm}
		\begin{tabular}{|c|c|c|c|c|c|c|c|c|c|c|c|} \hline
			Model & \multicolumn{5}{|c|}{Goodness-of-fit test $p$-value}&\multicolumn{6}{|c|}{Coverage rate}\\ \hline
			&T1  & T2 & T3 & T4 & T5& 0.05 & 0.1 & 0.25 & 0.75 & 0.9& 0.95\\
			\hline\hline 
			\multicolumn{12}{|c|}{CU Price REGIME 1} \\ \hline 
			Gaussian	&	$0.2510$ & $0.1330$ & $0.0730$ & $0.0590$ & $0.0860$ &0.0478&	0.0955&	0.2119&	0.7582&	0.9045&	0.9552
			\\\hline 
			$\alpha$-stable	&		$0.3120$ & $0.2990$ & $0.2730$ & $0.2050$ & $0.3800$&0.0507&	0.0985&	0.2299&	0.7493&	0.8985&	0.9463
			\\\hline \hline
			\multicolumn{12}{|c|}{USDPLN REGIME 1} \\ \hline 
			Gaussian	&		$0.0010$ & $0.0000$ & $0.0000$ & $0.0000$ & $0.0000$ &0.0269	&0.0627&	0.2328&	0.7940&	0.9224&	0.9403
			\\\hline 
			$\alpha$-stable&	$0.7260$ & $0.9530$ & $0.8960$ & $0.9510$ & $0.4250$ &0.0627	&0.1015&	0.2657&	0.7701&	0.9104&	0.9493
			\\\hline \hline
			\multicolumn{12}{|c|}{CU Price REGIME 2} \\ \hline 
			Gaussian	&	$ 0.0490$ & $0.0230$ & $0.0030$ & $0.0020$ & $0.0010$ &0.0496&	0.0827&	0.2151&	0.7920&	0.9078&	0.9527
			\\\hline 
			$\alpha$-stable&	$0.9970$ & $0.9750$ & $0.9820$ & $0.9630$ & $0.9720$& 0.0520&	0.0946&	0.2435&	0.7754&	0.8889&	0.9338
			\\\hline \hline
			\multicolumn{12}{|c|}{USDPLN REGIME 2} \\ \hline 
			Gaussian	&		$0.0000$ & $0.0010$ & $0.0020$ & $0.0000$ & $0.0000$&0.0449&	0.0851&	0.2246&	0.7801&	0.9243&	0.9504
			\\\hline 
			$\alpha$-stable&$0.7630$ & $0.6410$ & $0.4050$ & $0.4150$ & $0.3470$  &0.0496&	0.1135&	0.2364&	0.7376&	0.9031&	0.9456
			\\\hline 
		\end{tabular}
		\label{tab:table3} 
	\end{table*}
	
	\begin{table}[htbp] 
		\centering 
		\caption{{The parameters of the $\alpha$-stable distribution estimated for the residual time series corresponding to the VAR(1) model with independent components.}}\vspace{0.1cm}
		\begin{tabular}{|c|c|c|c|} \hline
			$\alpha$  & $\sigma$ & $\beta$ & $\mu$\\
			\hline\hline 
			\multicolumn{4}{|c|}{CU Price REGIME 1} \\ \hline 
			$1.9233$ & $0.0236$ & $-0.5953$ & $0.0007$ \\ \hline 
			\multicolumn{4}{|c|}{USDPLN REGIME 1} \\ \hline 
			$1.7214$ & $0.0114$ & $1.0000$ & $0.0016$ \\ \hline 
			\multicolumn{4}{|c|}{CU Price REGIME 2} \\ \hline 
			$1.8314$ & $0.0120$ & $-0.4291$ & $-0.0005$ \\ \hline 
			\multicolumn{4}{|c|}{USDPLN REGIME 2} \\ \hline 
			$1.8411$ & $0.0073$ & $-0.0280$ & $0.0000$\\\hline
		\end{tabular}
		\label{tab:table4} 
	\end{table}
	
	\subsection{Modelling of the copper price in PLN - the comparative study}
\begin{figure*}[t]
		\begin{center}
			\includegraphics[scale=0.6]{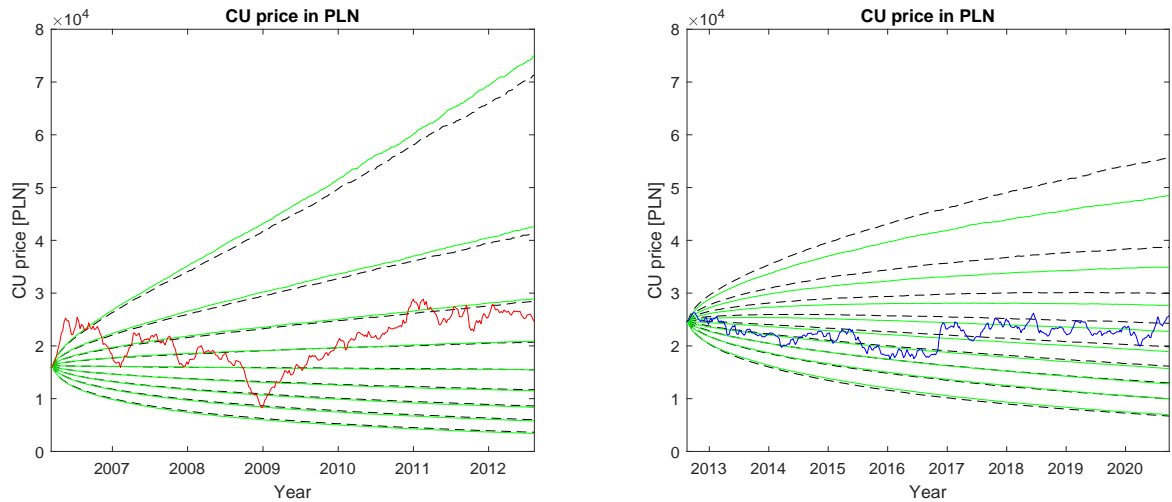}
			\caption{The copper (Cu) price in PLN in both analyzed regimes (blue solid lines) together with the quantile lines for $10\%, 20\%,...,90\%$ confidence levels for the VAR(1) model (black dashed lines) and for the VAR(1) model with independent components (green solid lines). }
			\label{fig7}
		\end{center}
	\end{figure*}

	Based on the models fitted to the logarithmic returns of the market quotations of the copper price in USD and the USDPLN exchange rate, we also infer the dynamics of the copper price in PLN, which is the main risk factor in KGHM mining company. To this end, we simulate the trajectories of the copper price in USD and USDPLN exchange rate using the fitted two-dimensional $\alpha-$stable VAR(1) models with the parameters given in Eq. (\ref{eq:Theta_Segments}) and in Eq. (\ref{eq:Theta_Segments_onedim}), respectively, i.e., when the relationship between the assets is taken under consideration or not. Then, the trajectories of the copper price in PLN are obtained as a product of the basic variables. The simulated trajectories are further used to derive the distribution of the copper prices in PLN. The obtained distributions are plotted in the form of quantile lines in Fig. \ref{fig7} for both regimes and both models. The calculations were based on $100000$ simulated trajectories.
	In the first regime, the copper price in PLN probability distribution is skewed to the upside reflecting the higher volatility of return rates both for the copper price and the USDPLN exchange rate and theoretically unlimited growth potential for the value of the assets in the extraordinary market environment. Using the $\alpha-$stable VAR(1) model taking into account the relationship between assets helps to narrow down the quantile lines, which should be supportive from the market risk management point of view, however, the improvement is not substantial. Regarding the second analyzed regime, the probability distribution for the copper price in PLN is more symmetric and narrower than in the first period. The $\alpha-$stable VAR(1) model with dependent factors does not show narrower quantile lines than in the case of no relation between both coordinates. Hence, for more stable market conditions the relationship between these two factors does not have to lead to narrowing price distributions. In a more stable situation on the market, often specific events, related exclusively to copper or USDPLN exchange rate drive their prices, which can create some volatility of the assets with harder to capture and modelling relations. In such circumstances, the two-dimensional model that assumes no dependence between components could give a similar outcome to the general VAR(1) model, when their relationship is taken into consideration.
	
	
	
	\section{Discussion and conclusions}
	
	The main purpose of the presented analysis was focused on the proper understanding of the dynamics of the analyzed historical data corresponding to the copper price in USD and USDPLN exchange rate and its implications for the copper price in PLN modelling. {The analysis was based on two specific risk factors, since they are the most important for the KGHM mining company, which is selling copper in USD and costs are incurred in PLN. For the company this can help,  to properly evaluate the potential downside market risk and to optimise the hedging instruments used for the risk management}. 
	
	We have proposed a two-dimensional VAR model with the $\alpha-$stable distribution that reflects the changing dependence structure of the analyzed assets. Although the negative dependence between the copper price and the exchange rate can not be simply explained by the market fundamentals, it is apparent in the market data, especially when there is an extraordinary black-swan event, such as the financial crisis of 2008. Omitting this issue in the construction of risk management strategies might result in suboptimal hedging schemes. On the other hand, the overestimation of the strength of this relation might cause large losses due to insufficient risk handling. Hence, allowing for a changing over time relation between two assets that reflects the actual market behaviour, incorporated in the proposed approach, might be utilized in developing more efficient risk management tools.
	
	We have started with measuring the correlation coefficients between the copper price in USD and the USDPLN exchange rate. The results clearly indicate that they differ substantially through time, regardless of the method used for the calculation (Pearson, Spearman rank, or Kendall rank correlations). Basing on this, the HMM approach was applied for the separation of two periods when the data behave differently. {The regimes were identified based on the $\alpha$-stable distribution assumption and possible changes of its parameters. The regime change was also reflected in two levels of the correlation coefficients.} The first period, around the financial crisis, has been characterized by higher volatility, excessive returns, and a negative correlation between the copper price in USD and the USDPLN exchange rate. The second period represents a more stable situation in the market. From the market risk perspective, both regimes are interesting, but taking into account the excessive risk for market participants which is related to higher volatility, analysis of the first regime is more important. 
	{One can notice that the correlation between two assets can substantially change through time and extraordinary price movements can occur in the foreseen moment. Moreover, the scale of the price is difficult to forecast. This implies that any hedging strategy or instruments used should be verified and constantly  monitored as ignoring the structural changes may lead to suboptimal asset allocation and hedging strategy mismatch. This is especially important when the perspectives of company management activities, including risk management, are focused on the mid- and long-term, like in the mining business. }
	
	Finally, we have estimated a two-dimensional vector autoregressive (VAR(1)) model with the $\alpha-$stable distribution, taking into account {a possible change of the scale and tail parameters} of the analyzed return rates for both assets. 
	{It should be highlighted, that the model incorporates dependence between the components, which reflects the single assets values changing over time. Moreover, incorporating the $\alpha-$stable distribution generalizes the standard Gaussian VAR model.} 
	For a comparison, we have also applied an approach with no relation between the assets, i.e the VAR(1) model with independent components.
	
	The comparative analysis of the results obtained with these two approaches leads to very interesting conclusions. On the one hand, the application of the $\alpha-$stable-based models seems to be more adequate for the analyzed data than the Gaussian-based approach. The copper prices and USDPLN exchange rates evidently exhibit heavy-tail behaviour. {The fit of the $\alpha$-stable VAR(1) model was confirmed by the residuals analysis as well as the coverage rates.} On the other hand, the application of the VAR(1) model allows for taking into account the relationship between the data that in general seems to be more appropriate than the approach when they are independent. From our analysis, we can conclude that in the first regime, taking into account the higher correlation between the risk factors in modelling does not lead to significantly narrower price distributions. From the market risk management perspective, this may imply that in the regime with the higher volatility, it is hard to control fat tails even taking into account the higher negative correlation between the assets.
	
	The careful investigation of the real data and the proper selection of the used methods enable building more adequate forecasts, especially for stress test scenarios. Such forecasts can be useful in analyzing liquidity risk, on the downside, as well as the potential impact of the royalty tax, on the upside, which is the power function of the copper price in PLN. 
	
	As the natural implication of the model fitted to assets traded on the market, we have derived also the dynamics of the copper price in PLN, which is not a traded asset but is crucial for the KGHM company risk exposure. Since for various international companies, the risk factors are given rather in the national than commodities market currency, the approach is universal and can be used in different market contexts, like mining or oil companies, but also other commodities involved in the global trading system.  {The analyzed model can be extended for other market risk factors of the KGHM and one can  conduct a similar analysis in a multi-dimensional way.}
	
	The critical aspect that should be further analyzed in the context of forecasting probability distributions is a proper detection of the moment when the regime switches. The ability to forecast such a moment with some advance would substantially increase the potential of using the presented approach in practice. We have shown that neglecting the regime change might cause large errors in the market predictions and has a significant impact on the efficient risk management tools.  
	
	One aspect we find interesting which could potentially improve further quality of the probability distribution of the copper price in PLN, is to include the inflation factor, which could also help to better define probability distributions projected for the unknown future. The cost of production curve among copper mining producers is one of the fundamental factors, which in some way limit downside potential for the price. In general, mining cost inflation, due to the costs related to the exploitation of more difficult mines and projects, is even higher than the most commonly used consumer or producer inflation indices. Therefore, taking into account inflation may improve the modelling approach.
	
	\section*{Acknowledgements}
	The work of A.W. was supported by National Center of Science under Opus Grant 2020/37/B/HS4/00120 "Market risk model identification and validation using novel statistical, probabilistic, and machine learning tools". J.J. acknowledges a support of NCN Sonata Grant No. 2019/35/D/HS4/00369.
	
	\bibliography{mybibliography}

	\section*{Appendix - Additional figures}
	
	\begin{figure}[ht!]
		\begin{center}
			\includegraphics[scale=0.5]{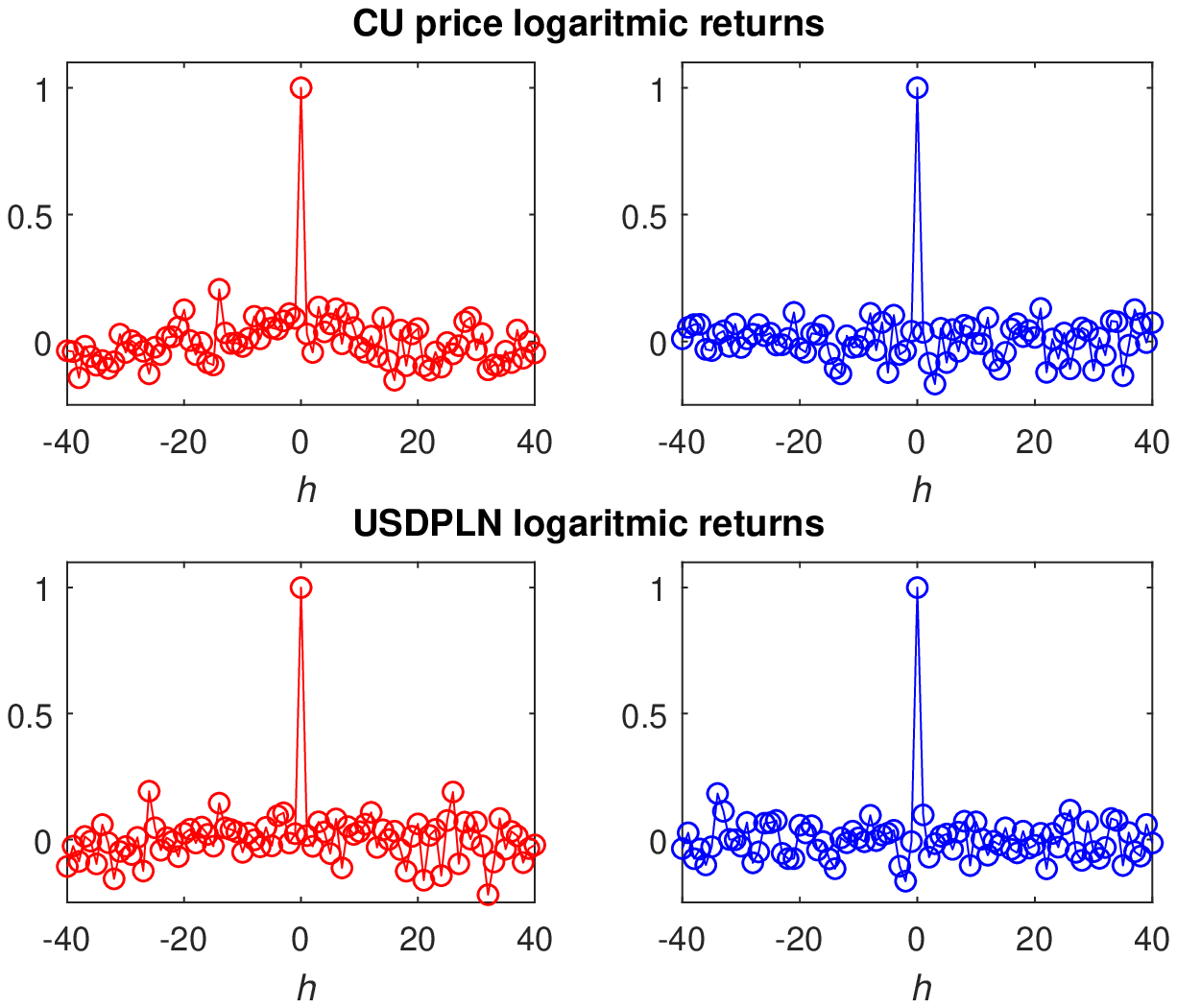}
			\caption{The normalized auto-covariation functions of the residuals corresponding to the VAR(1) models applied to the logarithmic returns of the weekly copper price in USD and USDPLN exchange rate for the data from regime 1 (left panels)  and regime 2 (right panels). }
			\label{fig_autocv_two}
		\end{center}
	\end{figure}
	
	
	\begin{figure}[ht!]
		\begin{center}
			\includegraphics[scale=0.5]{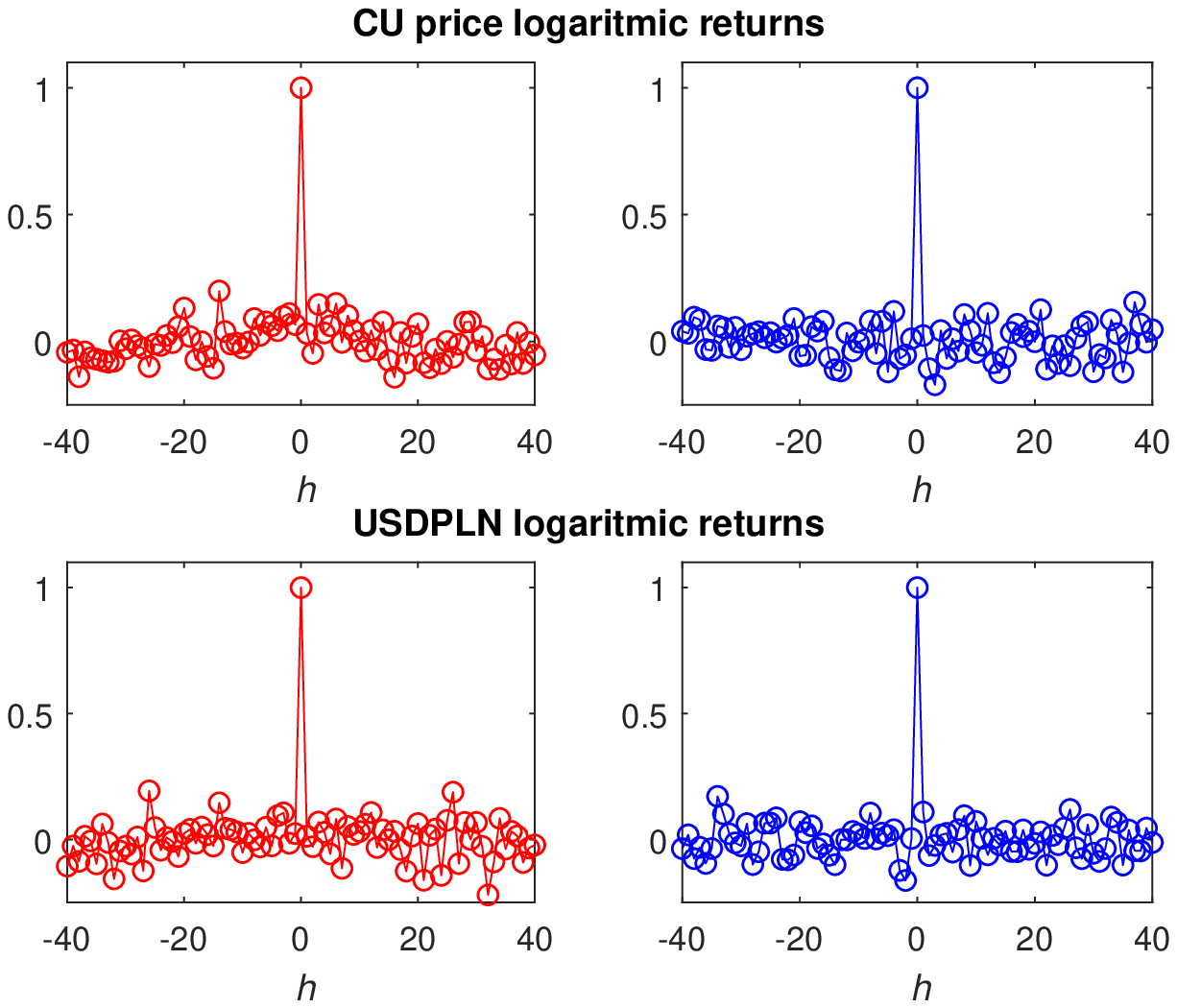}
			\caption{{The normalized auto-covariation functions of the residuals corresponding to the VAR(1) model with independent components applied to the logarithmic returns of the weekly copper price in USD and USDPLN exchange rate for the data from regime 1 (left panel)  and regime 2 (right panel).} }
			\label{fig_autocv_one}
		\end{center}
	\end{figure}
	
	
\end{document}